\documentclass{appolb}
\usepackage{graphicx}
\newcommand{\nn}{\nonumber}
\newcommand{\be}{\begin{equation}}
\newcommand{\ee}{\end{equation}}
\newcommand{\bea}{\begin{eqnarray}}
\newcommand{\eea}{\end{eqnarray}}

\usepackage{subfig}
\usepackage{graphicx}
\begin{document}
\title{The q-Statistics and QCD Thermodynamics at LHC%
\thanks{Presented at XI Workshop on Particle Correlations and Femtoscopy by R. Sahoo (Raghunath.Sahoo@cern.ch)}%
}
\author{Trambak Bhattacharyya, Arvind Khuntia, Pragati Sahoo, Prakhar  Garg, Pooja Pareek, Raghunath Sahoo
\address{Discipline of Physics, School of Basic Sciences, Indian Institute of Technology Indore, Khandwa Road, Simrol, M.P - 452020,
India.}
\\
\vskip 1.5em
{ Jean Cleymans}
\address{UCT-CERN Research Centre and Department of Physics, University of Cape Town, Rondebosch 7701, South Africa}
}
\maketitle
\begin{abstract}
We perform a Taylor series expansion of Tsallis distribution by assuming the Tsallis parameter $q$ close to 1. The $q$ value shows the deviation of a system from a thermalised Boltzmann distribution. By taking up to first order in $(q-1)$, we derive an analytical result for Tsallis distribution including radial flow. Further, in the present work, we also study the speed of sound ($c_s$) as a  function of temperature using the non-extensive Tsallis statistics for different $q$ values and for different mass cut-offs.
\end{abstract}
  
\section{Introduction}
Tsallis distribution has been successful in explaining the transverse momentum spectra  observed at the Relativistic Heavy Ion Collider (RHIC) and at the Large Hadron Collider (LHC) with a non-extensive parameter $q$.  Here $q$ shows the deviation of  a system from a thermalised Boltzmann system. We incorporate the radial flow in the Tsallis statistics by using Taylor series expansion to get an analytical expression. It is believed that $c_s^2$ in the hadronic medium decreases to zero in the first order phase transition from hadronic phase to QGP phase. Therefore, the study of $c_s^2$ is important to locate the QCD phase boundary.

\section{Taylor expansion of Tsallis distribution in $(q-1)$}
Tsallis non-extensive distribution function is given by,
\begin{equation}
f = \left[1 + (q-1)\frac{E-\mu}{T}\right]^{-\frac{1}{q-1}} .
\label{tsallis}
\end{equation}
 One obtains the relevant thermodynamical quantities by using $f^q$\cite {Bhattacharyya:2015hya}. Assuming the non-extensive parameter $q$ close to 1, Tsallis distribution function, $f^q, $ can be expanded in Taylor series in $(q-1)$, which is given by,
\begin{eqnarray}
&&\left[1+(q-1)\frac{E-\mu}{T}\right]^{-\frac{q}{q-1}}\nonumber\\
&\simeq& \mathrm{e}^{-\frac{E-\mu}{T}}\left\{ 1 +  
(q-1)\frac{1}{2}\frac{E-\mu}{T}\left( -2 + \frac{E-\mu}{T}\right)\right.+\frac{(q-1)^2}{2!}\nonumber\\
&&\frac{1}{12}\left[\frac{E-\mu}{T}\right]^2
\left[ 24 - 20\frac{E-\mu}{T} +3\left( \frac{E-\mu}{T}\right)^2\right]+ \mathcal{O}\left\{(q-1)^3\right\}+\left. . \right\}
\label{taylor}
\end{eqnarray}
This result is used to calculate the momentum spectra~\cite{Bhattacharyya:2015hya}. 
In Fig. 1 we show fits to the
normalized differential $\pi^-$ yields 
in $(0-5)\%$ Pb+Pb collisions  at $\sqrt{s_{\rm NN}} $ = 2.76 TeV 
with the Tsallis (solid line) 
and Boltzmann distributions (dashed line). Also, fits with the Tsallis distribution keeping terms to first order
(dash-dotted line) and second order in $(q-1)$ (dotted line) are shown in Fig. 1.
In Fig. 2  the fits are with the Tsallis distribution including flow  keeping terms to first  order in $(q-1)$ (dashed line). The flow velocity is fixed at $v = 0.609$, with $T = 146 $ MeV, $q = 1.030$ and the radius of the volume is $R = 29.8$ fm. The lower part of the figures show the deviation from data points. 
For truncation of the series expansion up to the first order with pure Tsallis distribution, one needs to satisfy two
conditions~\cite{Bhattacharyya:2015hya},  i.e.

$|1-q|\frac{E}{T}<1$
and
$q|1-q|\left(\frac{E}{T}\right)^2<2$.

\section{Momentum spectra including radial flow in $(q-1)$}
In order to incorporate the effect of radial flow on the momentum spectra and to see how it could improve to explain the momentum spectra obtained in Pb+Pb collisions, we have included a constant flow velocity, $v$, and the invariant yield is given by, 

\begin{eqnarray}
&&\frac{1}{p_T}\frac{dN}{dp_Tdy} = \frac{gV}{(2\pi)^2} \biggl\{ 2 T [ r I_0(s) K_1(r) - s I_1(s) K_0(r) ]\nn\\
&&-(q-1) T r^2  I_0(s) [K_0(r)+K_2(r)] + 4(q-1)~T rs I_1(s) K_1(r)\nn\\
&&-(q-1)Ts^2 K_0(r)[I_0(s)+I_2(s)] + \frac{(q-1)}{4}T r^3 I_0(s) [K_3(r)+3K_1(r)]\nn\\
&&-\frac{3(q-1)}{2}T r^2 s [K_2(r)+K_0(r)] I_1(s) + \frac{3(q-1)}{2} T s^2 r [I_0(s)+ \nn\\
&&I_2(s)] K_1(r)-\left.\frac{(q-1)}{4}T s^3 [I_3(s)+3I_1(s)] K_0(r)\right\}
\label{tsallisflow}
\end{eqnarray}
$r\equiv\frac{\gamma m_T}{T}$ , $s\equiv\frac{\gamma v p_T}{T}$, $I_n(s)$ and $K_n(r)$ are  the modified Bessel functions of the first
and second kind. 

\begin{figure}[ht!] \label{ fig7} 
  \begin{minipage}[b]{0.47\linewidth}
    \includegraphics[width=.82\linewidth]{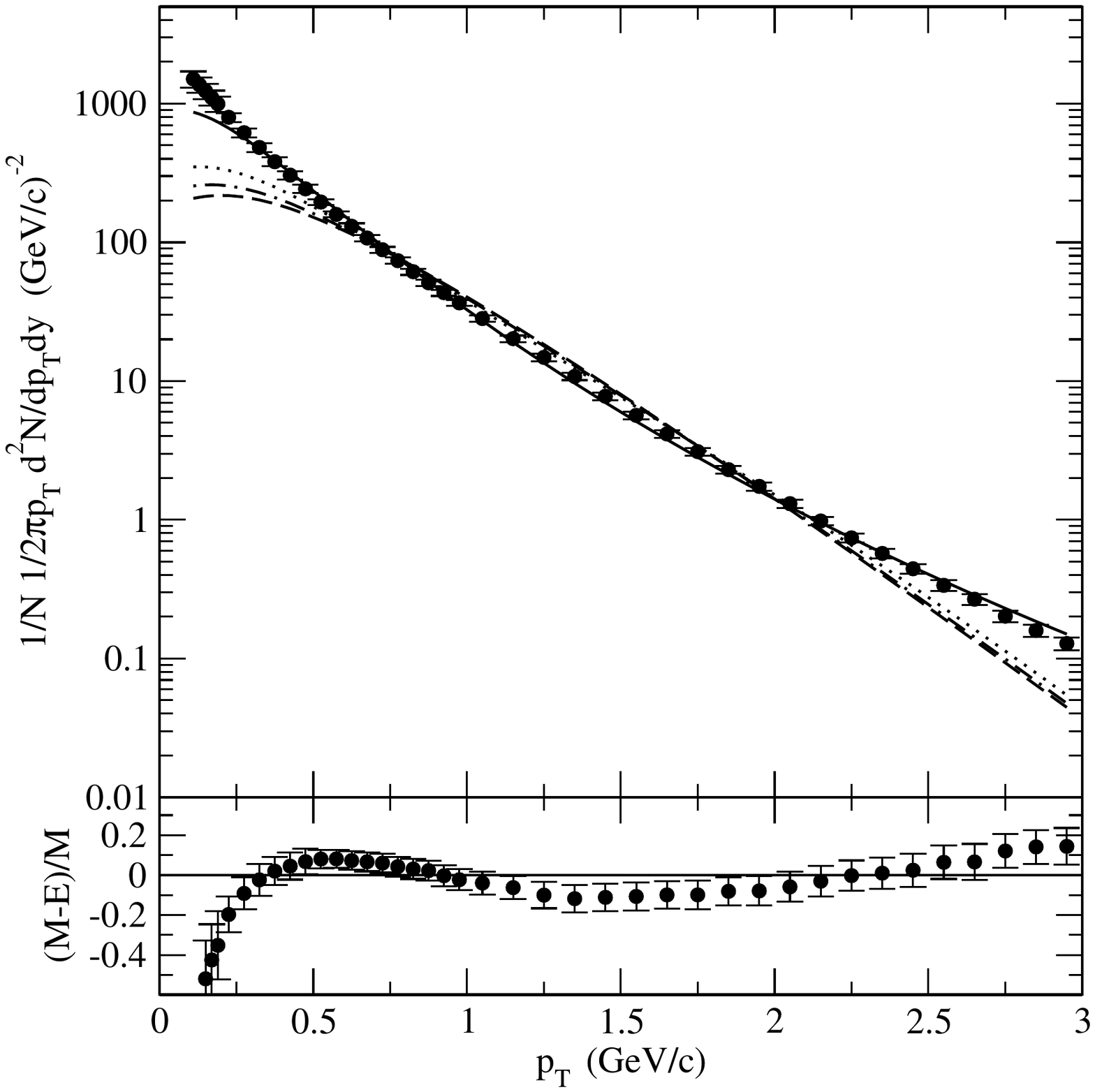} 
    \caption{Fits to the normalized differential $\pi^-$
yields in  $(0-5)\%$ Pb+Pb collisions at
$\sqrt{s_{\rm NN}}$ = 2.76 TeV~\cite{Bhattacharyya:2015hya} with the Tsallis distribution without radial-flow. } 
  \end{minipage} 
  \begin{minipage}[b]{0.47\linewidth}
    \includegraphics[width=.82\linewidth]{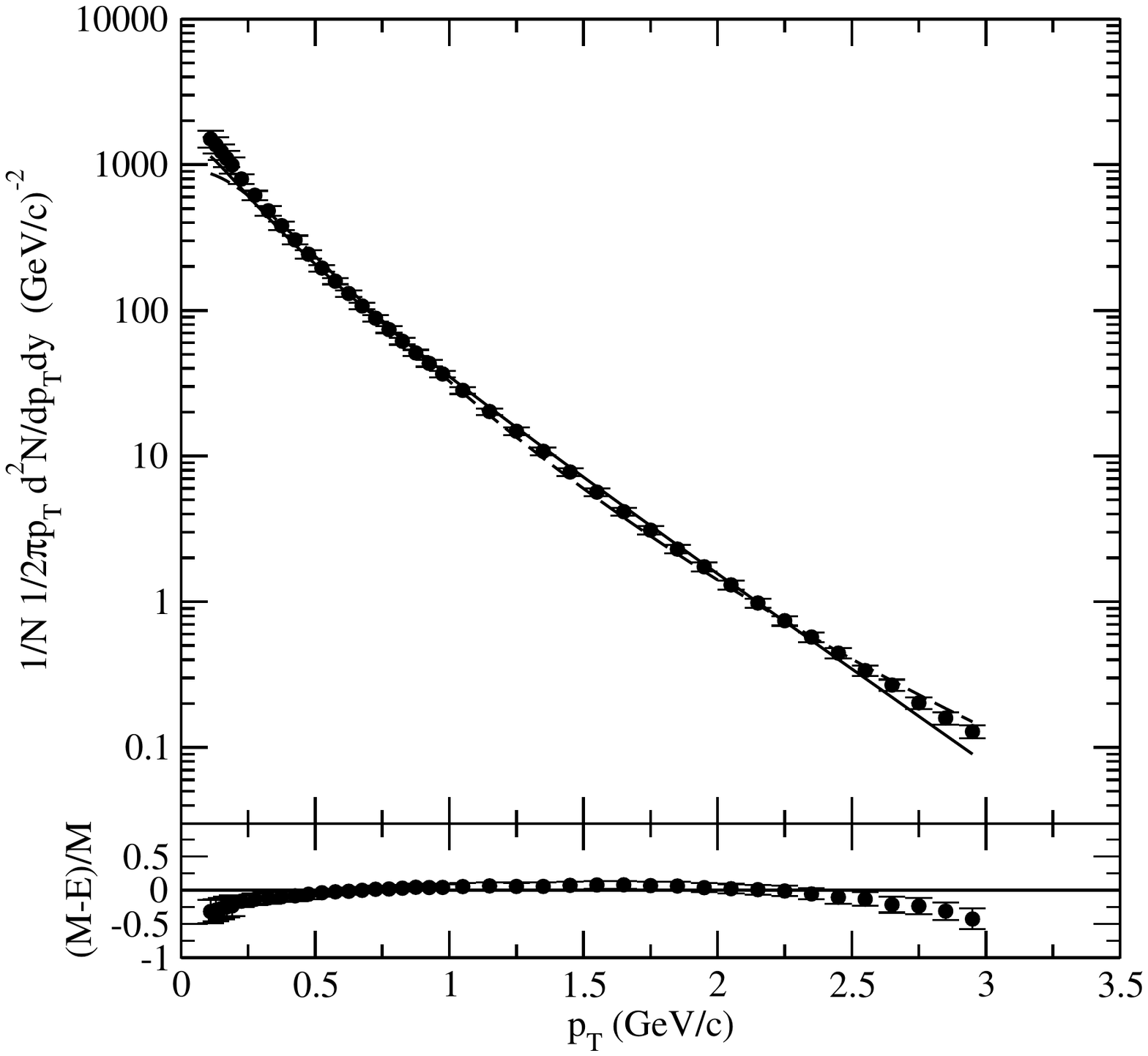} 
    \caption{ Fits to the normalized differential $\pi^-$
yields in  $(0-5)\%$ Pb+Pb collisions at
$\sqrt{s_{\rm NN}}$ = 2.76 TeV~\cite{Bhattacharyya:2015hya} with the Tsallis distribution with radial-flow (dashed line).} 
  \end{minipage} 
  \begin{minipage}[b]{0.45\linewidth}
    \includegraphics[width=.82\linewidth]{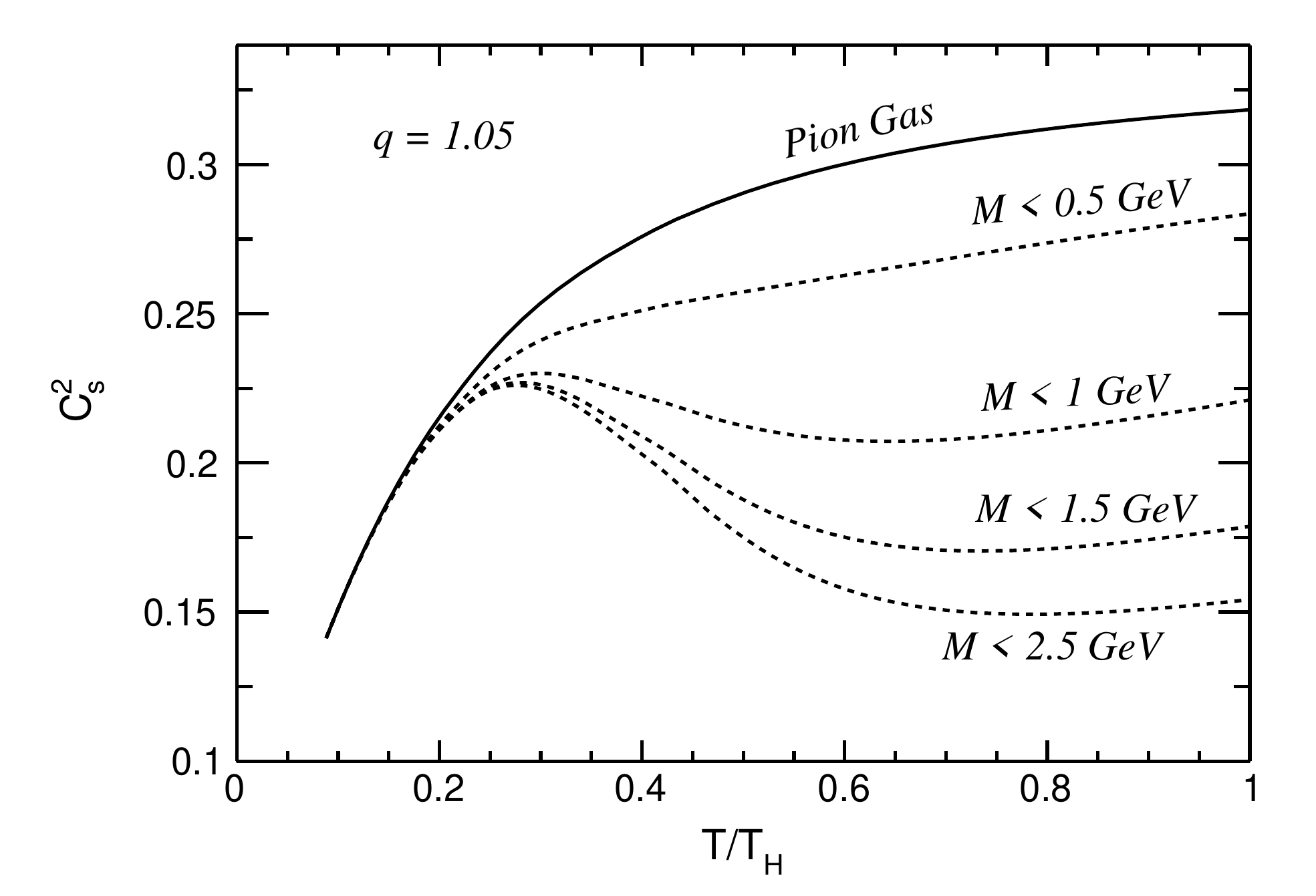} 
    \caption{The $c_s^2$ as a function of temperature for $q$=1.05~\cite{Khuntia:2016ikm}.} 
  \end{minipage}
  \hfill
  \begin{minipage}[b]{0.45\linewidth}
    \includegraphics[width=.85\linewidth]{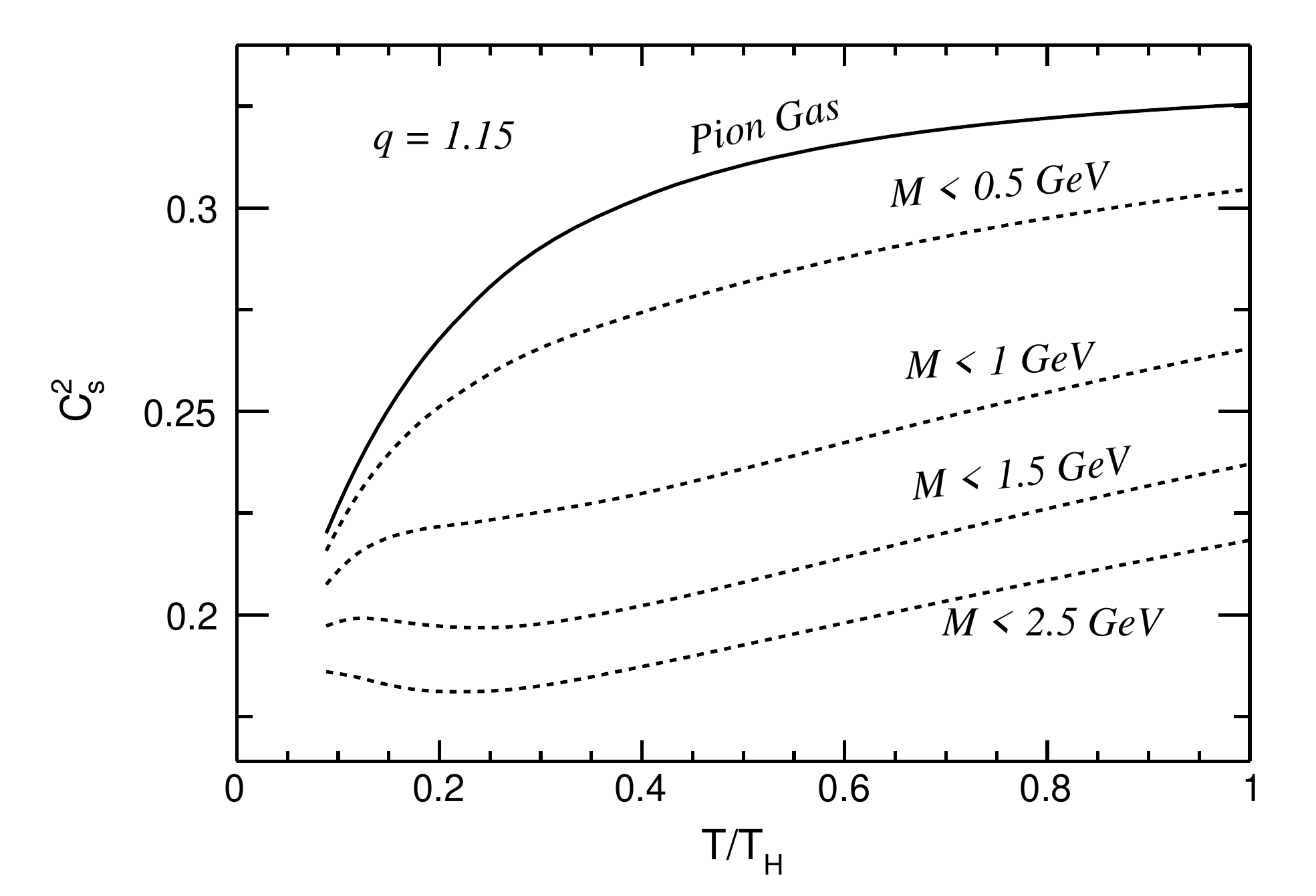} 
    \caption{The $c_s^2$ as a function of temperature for $q$=1.15~\cite{Khuntia:2016ikm}.} 
  \end{minipage} 
\end{figure}
Fig. 2 shows that our model explains experimental data very well in the intermediate $p_T$ region (0.5-2.5 GeV/c) but shows slight deviation in low and high $p_T$ regions.
 


\section{Speed of Sound in a Physical Hadron Resonance Gas}
For a physical hadron resonance gas, we use Tsallis form of Fermi-Dirac (FD) and Bose-Einstein (BE) statistics according to particle species \cite {Khuntia:2016ikm}, which is given by,
\begin{equation}
f_T(E) \equiv 
\frac{1}{\mathrm{\exp}_q\left(\frac{E-\mu}{T}\right) \pm1} .
\label{tsallis-fd}
\end{equation}
the $\pm$ sign in the denominator stands for FD and BE respectively.
 In the limit $q \rightarrow 1$,  exp$_q$ reduces to the standard exponential. The square of speed of sound is the change in pressure with change in energy density, which is given by
\begin{equation} 
c_s^2(T) = \left(\frac{\partial P}{\partial\epsilon}\right)_V ~=~ \frac{s(T)}{C_V(T)},
\label{velOfSound}
\end{equation}
where, s is the entropy and $C_V$ is the specific heat at constant volume.  $c_s^2$ is plotted as a function of temperature in Fig. 3 and Fig. 4, taking different mass cut-offs for $q$=1.05 and $q$=1.15.

\section{Summary}
Taylor series expansion of Tsallis distribution in $(q-1)$ is used to introduce the radial flow in an analytic way in addition to the study of the degree of deviation from equilibrium Boltzmann distribution. The $p_T$ spectra for 0-5 $\%$ central Pb+Pb collisions at 2.76 TeV is well described with inclusion of radial flow up to first order in $(q-1)$. The speed of sound for physical resonance gas has  been calculated for different non-extensive parameters $q$. Further, taking higher $q$ values seems to increase the speed of sound near the limiting temperature ($T_H$) as compared to the extensive Boltzmann statistics. For higher $q$ values the critical behaviour of speed of sound occurs earlier as compared to extensive statistics, which indicates an earlier phase transition in non-extensive statistics. This seems to be a function of degree of deviation from equilibrium statistics.


\end{document}